\documentstyle[preprint, aps]{revtex}

\begin{document}
\title{Reversed Dark Resonance in Rb Atom Excited by a Diode Laser}
\author{Janis Alnis, Marcis Auzinsh\thanks{%
Corresponding author, Fax +371-7820113, e-mail mauzins@latnet.lv}}
\address{Department of Physics, University of Latvia, 19 Rainis boulevard, Riga,\\
LV-1586, Latvia}
\date{\today}
\maketitle

\begin{abstract}
Origin of recently discovered reversed (opposite sign) dark resonances was
explained theoretically and verified experimentally. It is shown that the
reason for these resonances is a specific optical pumping of ground state
level in a transition when ground state angular momentum is smaller than the
excited state momentum.
\end{abstract}

\section{Introduction}

Coherent population trapping was discovered in the interaction of sodium
atoms with a laser field in 1976. \cite{moi1}. Due to this effect a
substantial part of population, because of destructive quantum interference
between different excitation pathways, is trapped in a coherent
superposition of ground state sublevels -- dark states. With a coherent
population trapping are associated dark resonances when due to this effect
absorption and as a result fluorescence from atoms decreases, but intensity
of the transmitted light increases when part of the atomic population is
trapped in dark states. If in addition to the optical excitation an external
magnetic field is applied, it can destroy coherence between ground state
sublevels and return trapped population into absorbing states and, as a
result, increase absorption and fluorescence, but decrease a transmitted
light. A review of applications of dark resonances was published some years
ago by Arimondo \cite{ar1}. Coherence in an atomic ground state attracted
substantial attention in connection with lasing without inversion \cite{sc1}%
, magnetometry \cite{sc2} and laser cooling \cite{as1}. As a result dark
resonances recently are studied in detail, including open systems \cite{re1}
and systems with losses \cite{re2}. In course of these studies a new and
unexpected phenomenon was observed by authors of \cite{da1}. In this study D$%
_{2}$ line of $^{85}$Rb atoms was excited by a diode laser. Radiation was
tuned to the absorption from optically resolved ground state hyperfine level 
$F_{g}=3$ originating from atomic $5$S$_{1/2}$ state. The final state of the
transition was Rb $5$P$_{3/2}$ excited state. Hyperfine components of this
level was not resolved and all allowed in a dipole transition excited state
hyperfine levels with quantum numbers $F_{e}=2,3,4$ were excited. In
contrary to the usual dark resonance signal when in the absence of the
magnetic field one can observe increased transmittance and decreased
fluorescence intensity authors observed opposite effect --- decreased
transmittance which increases with magnetic field applied and an increased
fluorescence intensity which decreased when field was applied. Authors of
the paper \cite{da1} write that the physical reason of this effect {\it %
remains still unclear}. They suppose that one of the reasons for the peak in
the fluorescence could be the non-coupled states on the Zeeman sublevels of
the excited hyperfine levels. These non-coupled states, as it is supposed in 
\cite{da1}, inhibit the stimulated emission induced by the laser field. The
decrease of stimulated emission leads to an increase of the fluorescence.

In this letter we offer, in our opinion, very simple and strait forward
explanation of the origin of these ''reversed'' dark resonances and perform
experimental and numerical studies of them.

\section{Reversed resonance}

In a simple qualitative explanation traditional dark resonances can be
connected with a well known optical pumping phenomenon. Let us assume that
we excite atomic transition $F_{g}=2\longrightarrow F_{e}=1$ with a linearly
polarized light. Direction of the $z$ axis is chosen along the light
electric field vector ${\bf E}$. As a result $\pi $ absorption takes place
and transitions occur between ground and excited state magnetic sublevels
with $\Delta M=M_{g}-M_{e}=$ $0$, where $M_{g},M_{e}$ are magnetic quantum
numbers of the ground and excited states respectively, see Fig. \ref{fig1}.
According to this scheme absorption does not take place from ground state
magnetic sublevels with quantum number $M_{g}=\pm 2$, because for these
states there are no corresponding excited state magnetic sublevel with the
same magnetic quantum number value.

In the spontaneous decay dipole transitions from optically populated excited
state magnetic sublevels $M_{e}=\pm 1$ to the nonabsorptive ground state
sublevels $M_{g}=\pm 2$ are allowed. As a result, if relaxation in the
ground state is slow, in a steady state conditions substantial part of the
population will be optically pumped to the ground state sublevels with
quantum numbers $M_{g}=\pm 2$ and will be trapped there. As a result
traditional decrease of the absorption and fluorescence and increase of the
transmittance will be observed, because the population of absorbing ground
sate magnetic sublevels will be reduced.

If we now apply an external magnetic field in a direction perpendicular to
the $z$ axis, field will mix ground state sublevels effectively and will
return trapped population into the states from which absorption takes place.
As a result absorption and fluorescence will increase. This is a qualitative
explanation of the usual dark resonance.

A similar reasoning can be exploited to explain the ''reversed'' resonance
observed in \cite{da1} and in this paper. Let us assume that we excite with $%
\pi $ radiation atomic transition $F_{g}=1\longrightarrow F_{e}=2$. In this
case there are no ground state sublevels not involved in the absorption that
can trap atomic population. The actual relative transition rates
proportional to the squared respective Clebsch -- Gordan coefficients in
this system of sublevels are shown in Fig. \ref{fig2}. As one can see ground
state magnetic sublevel $M_{g}=0$ is the most absorbing - with highest
relative absorption rate. At the same time just to this sublevel intensively
with high rates decay all three excited state\ magnetic sublevels populated
by the light. One can expect that in a conditions of a steady state
excitation, as a result of interplay of absorption and decay rates,
population of the intensively absorbing ground state magnetic sublevel $%
M_{g}=0$ will be increased and, as a result, one can expect increased
absorption and fluorescence from this atom and decreased transmittance of
the resonant laser light.

If an external magnetic field is applied perpendicularly to $z$ axis it will
mix ground state magnetic sublevels and redistribute population between the
ground state magnetic sublevels. As a result population of intensively
absorbing magnetic sublevel $M_{g}=0$ will be decreased. At the same time
population of less absorbing magnetic sublevels $M_{g}=\pm 1$ will be
increased \ This means that the total absorption and fluorescence will be
decreased and transmittance will be increased. This means that reversed dark
resonance will be observed.

To prove this qualitative consideration let us solve balance equations for
the magnetic sublevel stationary population $n_{M_{g}}$ in the scheme shown
in Fig. \ref{fig2}. In a steady state conditions for the ground state
magnetic sublevels we will obtain 
\begin{eqnarray}
n_{-1} &=&\frac{6(6\Gamma +5\Gamma _{p})}{51\Gamma +100\Gamma _{p}}\overline{%
n_{g}},  \nonumber \\
n_{0} &=&\frac{9(9\Gamma +10\Gamma _{p})}{51\Gamma +100\Gamma _{p}}\overline{%
n_{g}},  \label{eq1} \\
n_{+1} &=&\frac{6(6\Gamma +5\Gamma _{p})}{51\Gamma +100\Gamma _{p}}\overline{%
n_{g}},  \nonumber
\end{eqnarray}
where $\Gamma $ is excited state relaxation rate, $\Gamma _{p}$ absorption
rate, and $\overline{n_{g}}$ ground state magnetic sublevel population in
absence of the radiation. In a condition when absorption is slow $\Gamma
_{p}\ll \Gamma $ --- weak absorption, we have 
\begin{eqnarray}
n_{-1} &\approx &\frac{36}{51}n_{g}\approx 0.706\overline{n_{g}},  \nonumber
\\
n_{0} &\approx &\frac{81}{51}n_{g}\approx 1.59\overline{n_{g}},  \label{eq2}
\\
n_{+1} &\approx &\frac{36}{51}n_{g}\approx 0.706\overline{n_{g}}.  \nonumber
\end{eqnarray}
If we now keep in mind absorption rates from different magnetic sublevels of
the ground state, see Fig. \ref{fig2}, and calculate the overall absorption
from such state and compare it with the absorption from the equally
populated magnetic sublevels (when magnetic field is applied) than we see an
increase in the absorption rate by a factor $18/17\approx 1.059$ or by
approximately $5.9\%$.

The same calculation can be performed for the transitions $%
F_{g}=2\longrightarrow F_{e}=3$ and $F_{g}=3\longrightarrow F_{e}=4$. For
these schemes in a similar way we will obtain even larger increase of the
absorption due to this specific optical pumping. The increase will be by a
factor $540/461\approx 1.17$ and $4004/3217\approx 1.24$ respectively. This
means that the described effect increases with increase of the quantum
numbers of involved levels.

Of course presented description is only qualitative, but in our opinion
gives a good idea what is happening when reversed dark resonances are
observed. To have a quantitative description of the phenomenon one must
solve equations for the density matrix for an open system with losses. We
will not do this in present paper. Instead a simple analysis will be carried
out.

An analysis of the probabilities of optical transitions originating from $%
F_{g}=3$ between excited hyperfine levels of the Rb atom show that levels $%
F_{e}=2,3,4$ are populated in the ratio ($5/18\approx 0.278):(35/36\approx
0.\allowbreak 972):(9/4=2.25)$. It means that a hyperfine transition leading
to the reversed dark resonances discussed above is most strongly excited.
For this scheme let us calculate a signal shape using a full density matrix
approach. We solved a rate equations for a density matrix, see \cite{au1},
Chapter 5, for $F_{g}=3\longleftrightarrow F_{e}=4$ transition. A broad line
approximation was used. It means that we assume that in a magnetic field all
magnetic sublevels are in equally good resonance with radiation. Secondly,
we assumed that at a magnetic field strength used in the experiment
(resonance width is less than 100 mG) hyperfine levels experience linear
Zeeman effect. No substantial hyperfine level mixing at applied field
strength takes place. Direct Rb atom magnetic sublevel splitting in a
magnetic field calculations and measurements prove that these assumption are
valid, see for example \cite{au2}. For signal simulation the following rate
constants were used. Excited state relaxation rate $\Gamma =3.8\times 10^{7}$
s$^{-1}$ \cite{sva1}, absorption rate $\Gamma _{p}=3\times 10^{6}$ s$^{-1}$,
ground state relaxation rate $\gamma =2\times 10^{5}$ s$^{-1}$ (mainly due
to collisions with the walls of the cell and fly-through the excitation
laser beam) . Lande factors $g_{g}=-0.3336$, $g_{e}=-0.5013$ were calculated
in a standard way from the atomic and nuclear data available in \cite{ari1}.
We suppose that a magnetic field is applied along $z$ axis. Laser light
excites $F_{g}=3\longrightarrow F_{e}=4$ transition and is linearly
polarized along $\ y$ axis. Intensity of the fluorescence with the same
polarization is calculated and the intensity of the transmitted beam is also
calculated as a function of the magnetic field. The results are presented in
a Fig. \ref{fig3}. They demonstrate well pronounced reverse resonances and
are in a very good qualitative agreement with the measurements obtained in 
\cite{da1}, see Fig. 5. there. Width of these resonances can be varied by
changing ground state Lande factor value, ground state relaxation rate and
absorption rate.

At week absorption $\Gamma _{p}\ll \Gamma $ the dark resonance width is
determined by a condition when ground state Larmor frequency is equal to the
ground state relaxation rate. For $F_{g}=3$ state of $^{85}$Rb atom at low
concentration it can be as narrow as $20-30$ mG. This is a width that was
actually observed in \cite{da1}.

Obtained signals in some sense are the same as the ground state Hanle effect
measured in atoms as well as in molecules in great extent, see. for example 
\cite{au1,au3}. In case of molecules also reversed structure in ground state
Hanle effect was observed. In case of molecules, when optical pumping takes
place in an open cycle and a total ground state population is substantially
reduced, this structure can be attributed to a high order coherence created
between ground state magnetic sublevels \cite{au4,au5}

\section{Experimental}

We performed measurements of these reversed resonances also in our
laboratory. In our experiment we use isotopically enriched rubidium (99 \%
of $^{85}$Rb) that is contained in a glass cell at room temperature to keep
atomic vapor concentration low and avoid reabsorption. Transition $5s$ $%
^{2}S_{1/2}$ to $5p$ $^{2}P_{3/2}$ at $780.2$ nm is excited using both
temperature- and current-stabilized single-mode diode laser (Sony SLD114VS)
with beam diameter $7$ mm. Absorption signal (transmitted light) is
monitored by a photodiode. As the laser is swept applying a ramp on a drive
current, two absorption peaks with half-width of about $600$ MHz separated
by $\sim 3$ GHz appear due to the $^{85}$Rb ground state hyperfine
structure. The excited state hyperfine structure is not resolved under
Doppler profile.

During the resonance measurements, the laser wavelength is stabilized on
absorption peak originating from ground sate hyperfine level $F_{g}=3$.

Helmholz coils are used to sweep the magnetic field over zero Gauss region,
the Earth magnetic field components are compensated.

Signal detected in a transmitted light is averaged over 64 cycles and the
result is presented in a Fig. \ref{fig4}, data points and Lorenz fitting,
curve 1. On the same figure a calculated signal, curve 2, is presented.
Model for this calculation is the same as in case on Fig. \ref{fig3}.
Amplitude of the experimental and calculated signals are in arbitrary units
and in figure are scaled in a way to make them easy to compare. For
theoretical curve the parameters are chosen to have values that maximally
reproduce our experimental conditions. The ground state relaxation rate $%
\gamma =v_{p}/r_{0}=0.07$ $\mu $s$^{-1}$ was chosen as a reciprocal time of
thermal motion of Rb atoms at room temperature with most probable velocity $%
v_{p}=0.24$ mm/$\mu s$ through the laser beam of radius $r_{0}=3.5$ mm \cite
{au1}. Absorption rate was chosen to be $\Gamma _{p}=1.5$ $\mu $s$^{-1}$.
Other parameters are as in Fig. \ref{fig3}. Agreement between calculated and
measured signal is remarkable. As far as our model does not account for
transitions to other hyperfine levels and so does not reproduce experiment
in full, we do not attempt to fit experimental points with theoretical
curve, nevertheless achieved agreement fully convinces us that the proposed
explanation of the reversed dark resonances is correct. To obtain
quantitative description of this signal in future one must take into account
in the numerical model all other hyperfine levels involved in the process.

\section{Acknowledgment}

One of us MA is thankful to Prof. Neil Shafer-Ray for fruitful discussions.
Financial support from Swedish Institute Visby program is greatly
acknowledged.

\begin{figure}[tbp]
\caption{{}Allowed dipole transition scheme for ground state optical pumping
in case of $\protect\pi $ absorption for $F_{g}=2\rightarrow F_{e}=1$.}
\label{fig1}
\end{figure}

\begin{figure}[tbp]
\caption{{}Transition scheme and rate constants for $\protect\pi $
absorption in case of $F_{g}=1\rightarrow F_{e}=2.$}
\label{fig2}
\end{figure}

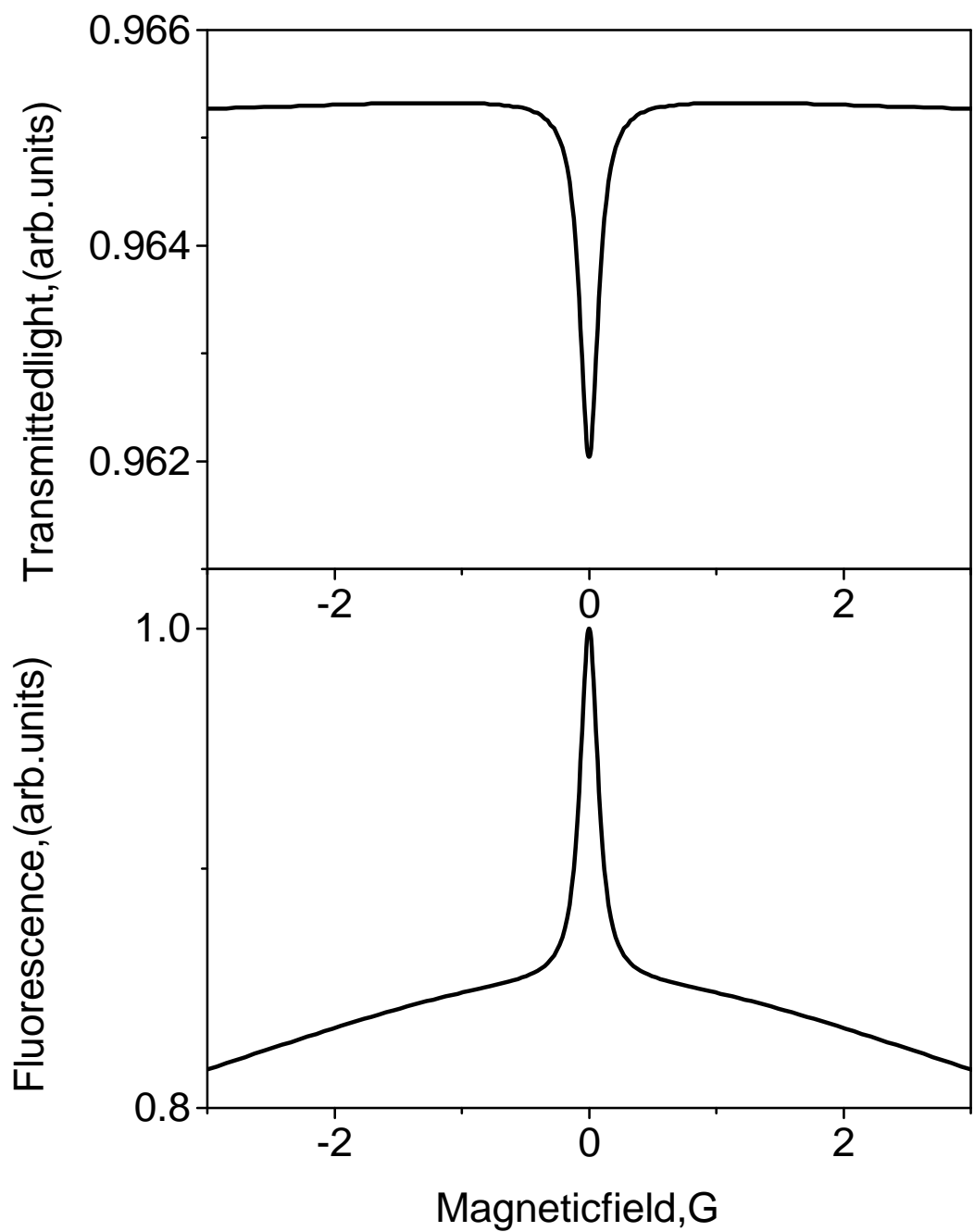
\begin{figure}[tbp]
\caption{{}Calculated intensity of fluorescence and transmitted light for
reverse dark resonance for $F_{g}=3\rightarrow F_{e}=4.$}
\label{fig3}
\end{figure}

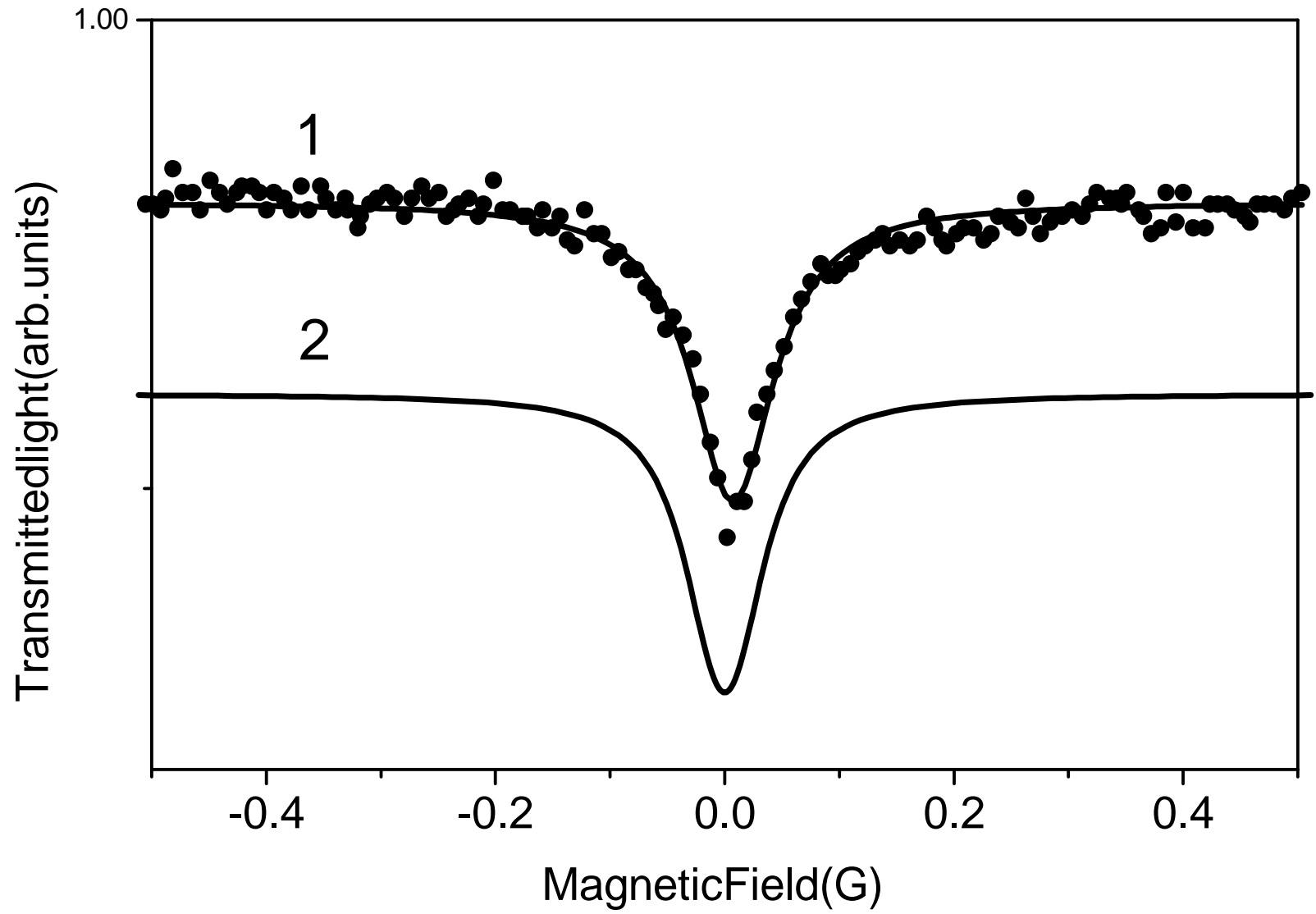
\begin{figure}[tbp]
\caption{{}Measured (points and Lorenz fitting --- curve 1) and calculated
(curve 2) reversed dark resonance in $^{85}$Rb.}
\label{fig4}
\end{figure}


\begin{references}
\bibitem{moi1}  G. Alzetta, A. Gozzini, L. Moi, and G. Orrioli, Novo Cimento
B {\bf 36}, (1976) 5

\bibitem{ar1}  E. Arimondo, Progr. Opt. {\bf 35} (1996) 257

\bibitem{sc1}  M. Scully, S-Y Zhu, A. Gavrielides, Phys. Rev. Lett. {\bf 62}
(1989) 2813

\bibitem{sc2}  M. Scully, M. Fleischhauer, Phys. Rev. Lett. {\bf 69} (1992)
1360

\bibitem{as1}  A. Aspect, E. Arimondo, R. Kaiser, N. Vansteenkiste, C.
Kohen-Tannoudji, Phys. Rev. Lett. {\bf 61} (1988) 826

\bibitem{re1}  Ferrucio Renzoni, Albrecht Lindner, and Ennio Arimondo, Phys.
Rev. A {\bf 60}, (1999) 450

\bibitem{re2}  F. Renzoni, W. Maichen, L. Windholz, and E. Arimondo, Phys.
Rev. A {\bf 55} (1997) 3710

\bibitem{da1}  Y. Dancheva, G. Alzetta, S. Cartalava, M. Taslakov, Ch.
Andreeva, Opt. Comm. {\bf 178} (2000) 103

\bibitem{au1}  M. Auzinsh, R. Ferber, {\it Optical Polarization of Molecules}%
, Cambridge University Press, Cambridge UK, 1995, 305

\bibitem{au2}  J. Alnis, M. Auzinsh, Phys. Rev. A, submitted

\bibitem{sva1}  G. Belin, S. Svanberg, Physica Scripta {\bf 4}, 269 (1971)

\bibitem{ari1}  E. Arimondo, M. Inguscio, P. Violino,{\it \ }Rev. Mod. Phys, 
{\bf 49}, 31 (1977)

\bibitem{au3}  M.P. Auzinsh, R.S. Ferber, Phys. Rev. A, {\bf 43}, 2374 (1991)

\bibitem{au4}  M.P. Auzinsh, R.S. Ferber, Sov. Phys. Usp. {\bf 33}, 833
(1990)

\bibitem{au5}  M.P. Auzinsh, R.S. Ferber, Opt. Spectrosc. (USSR), {\bf 55,}
674 (1983)
\end{references}
\end{document}